\documentclass[RNAAS]{aastex62}
\usepackage{times}

\begin{document}

\title{Some Turbulent Predictions for Parker Solar Probe}

\correspondingauthor{Steven R. Cranmer}
\email{steven.cranmer@colorado.edu}

\author[0000-0002-3699-3134]{Steven R. Cranmer}
\affiliation{Department of Astrophysical and Planetary Sciences,
Laboratory for Atmospheric and Space Physics,
University of Colorado, Boulder, CO 80309, USA}


\section{} 

From the the solar photosphere to the outer heliosphere, the Sun's
plasma properties are fluctuating with a broad range of temporal
and spatial scales.
In fact, a turbulent cascade of energy from large to small scales
is a frequently invoked explanation for heating the corona and
accelerating the solar wind.
NASA's {\em{Parker Solar Probe}} ({\em{PSP}}) is expected to
revolutionize our understanding of coronal heating and
magnetohydrodynamic (MHD) turbulence by performing in~situ sampling
closer to the Sun than any other prior space mission \citep{Fox16}.

This research note presents theoretical predictions for some
properties of MHD turbulence in the regions to be explored
by {\em{PSP.}}
These results are derived from a previously published
semiempirical model of coupled Alfv\'{e}nic and fast-mode
turbulence in the fast solar wind \citep{CvB12}.
This model contained predictions for the three-dimensional
wavenumber power spectra of incompressible Alfv\'{e}n waves
(which fluctuate transversely to the background magnetic field
and prefer to cascade to high perpendicular wavenumbers) and
fast-mode waves (which cascade isotropically in wavenumber space
and become compressible when propagating obliquely to the field)
as a function of radial distance from the Sun.

\begin{figure}[h!]
\begin{center}
\includegraphics[scale=0.79,angle=0]{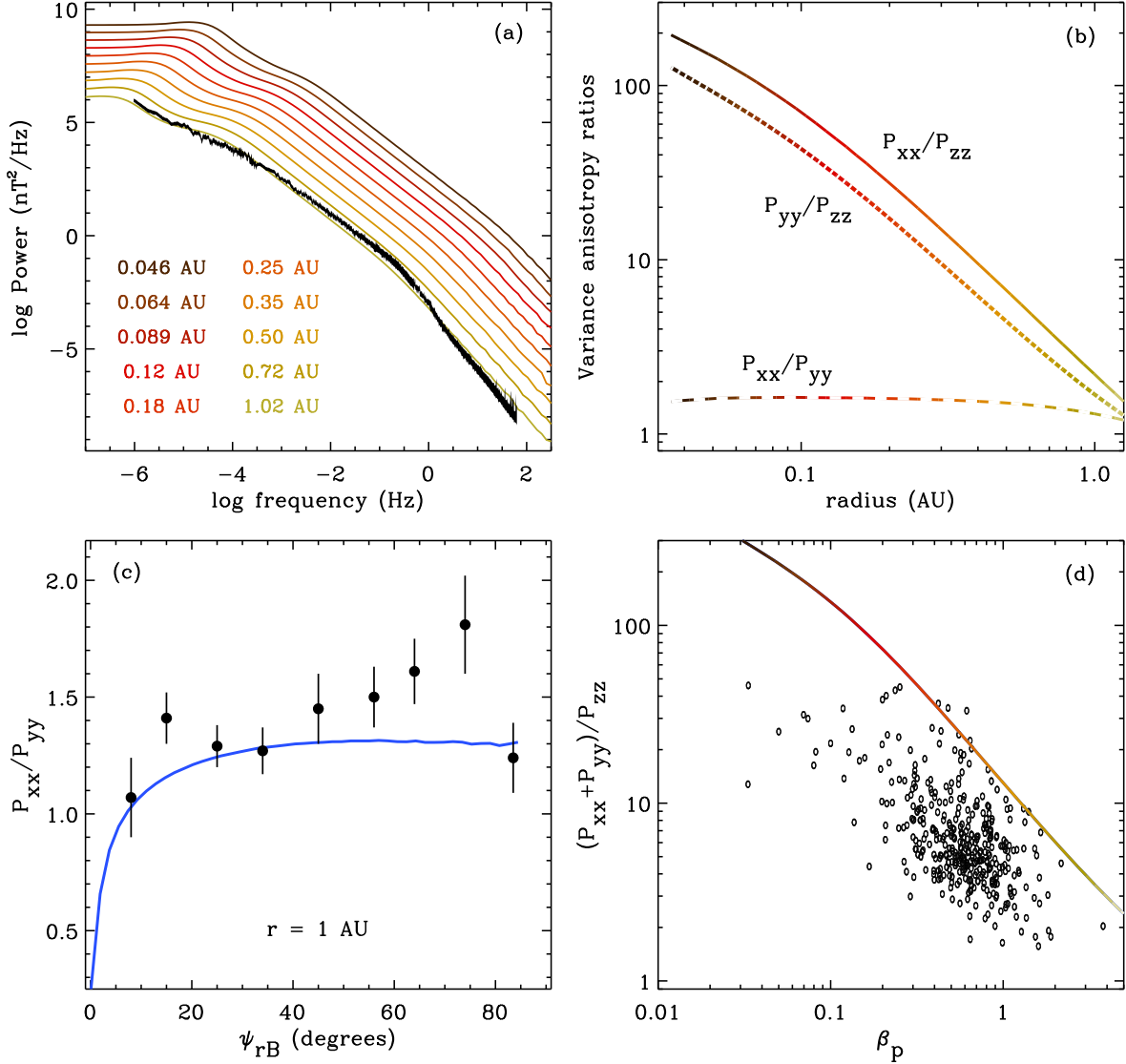}
\caption{(a) Spacecraft-frame power spectra of magnetic fluctuations
between 0.046 and 1.02 AU.
(b) Predicted variance anisotropy ratios versus distance.
(c) Dependence of $P_{xx}$:$P_{yy}$ on the field-flow angle at 1~AU.
(d) Dependence of total (perpendicular/parallel) variance anisotropy
on the proton plasma $\beta_p$ ratio.
The same color-scheme is used in panels (a), (b), and (d) for
the model data at different heliocentric distances.
See text for additional details.
Model data from all four panels are available as
``Data behind the Figure.''
\label{fig01}}
\end{center}
\end{figure}

Figure \ref{fig01}(a) shows spacecraft-frame frequency spectra
computed for the Alfv\'{e}nic magnetic fluctuations as a function
of radius.
For the purpose of this calculation, the background solar wind
model was changed from the original \citet{CvB12} polar field
line to one in the ecliptic plane (i.e., now including a mean
Parker spiral angle of $\sim${45\arcdeg} at 1 AU).
Multi-spacecraft measurements at 1 AU, presented originally by
\citet{Kiyani15}, are shown for comparison.
The lowest frequencies are dominated by anisotropic energy injection,
and the following outer-scale parameters from \citet{CvB12} were
chosen: $k_{0 \perp} = 10 / \lambda_{\perp}$ and
$k_{0 \parallel} = 0.01 k_{0 \perp}$.
Intermediate frequencies show the spacecraft-frame projection of an
inertial range dominated by wavenumber advection/diffusion that
obeys a so-called critical balance form of MHD turbulence \citep{GS95}.
The highest frequencies show the combined effects of
(1) linear wave damping from a full numerical solution of the
Vlasov-Maxwell dispersion relation, and (2) modified cascade due
to the kinetic Alfv\'{e}n wave (KAW) part of the dispersion relation.

Figure \ref{fig01}(b) shows the radial dependence of variance
anisotropy ratios of magnetic fluctuations (due to both Alfv\'{e}n
and fast modes) that were originally discussed by \citet{Belcher71}.
For Cartesian coordinates defined by the $z$ axis parallel to
the mean magnetic field, $x$ perpendicular to both $z$ and the
radial direction, and $y$ completing the orthogonal system,
\citet{Belcher71} found mutual ratios of order 5:4:1 for the power
variances $P_{xx}$:$P_{yy}$:$P_{zz}$ at 1~AU.
More recent measurements tend to show slightly lower values of both
$P_{xx}$ and $P_{yy}$ relative to $P_{zz}$ and similar ratios of
$P_{xx}$ to $P_{yy}$ \citep[see][]{Bieber96,MacBride10}.
The model results shown here were computed using the formalism given
by \citet{Wicks12}, and they used a mode coupling strength of
$\Phi = 10$ and a fast-mode power normalization of $0.3 U_{\rm F}$
\citep[for details, see][]{CvB12}.
In this model, the total amount of power in $P_{zz}$ depends only
on the fast-mode fluctuations, and the power in the two perpendicular
directions depends on both fast-mode and Alfv\'{e}n fluctuations.
The variance anisotropy is scale-dependent, and a representative
spacecraft-frame frequency of $10^{-2}$~Hz was chosen as being
reasonably comparable with existing measurements.

In Figure \ref{fig01}(c) the ratio of $P_{xx}$ to $P_{yy}$
at 1~AU is computed for a range of field-flow angles (i.e., the
angle between the $z$ axis and the radial direction, which is fixed
at {45\arcdeg} at 1~AU for the other panels) and it is compared
with data-points from Figure 1 of \citet{Bieber96}.
Figure \ref{fig01}(d) shows the dependence of the total variance
anisotropy ratio (i.e., $P_{xx}+P_{yy}$ to $P_{zz})$ on $\beta_p$,
the ratio of proton thermal pressure to magnetic pressure.
It is compared with {\em Helios} data-points from Figure 7 of
\citet{MacBride10}.

Despite some agreement in trends, it is nevertheless the case that
the model predictions do not perfectly match the existing observational
data.
The input parameters of the \citet{CvB12} model were not finely tuned
in order to match the in-ecliptic conditions to be seen by {\em{PSP.}}
The primary reason for this research note is to show how
straightforward it can be to extract useful predictions from existing
theoretical models about measurable quantities that were not even
considered when creating the models initially.
The variance anisotropy, for example, may be an important quantity
for distinguishing between different theoretical models for coronal
heating and solar wind acceleration.

\acknowledgments

Preliminary versions of these results were presented at the
{\em{PSP}} Science Working Group meeting in October 2017.
Hearty thanks go out to the team members who planned, built,
and launched {\em{PSP}} on its journey into the corona.

\end{document}